\documentclass[preprint, showpacs, preprintnumbers, amsmath, amssymb]{revtex4}
\usepackage{graphicx}
\usepackage{dcolumn}
\usepackage{bm}

\begin{document}

\title{Knotted Solitons in a Charged Two-Condensate Bose System}
\author{Yishi Duan}
\author{Xinhui Zhang}
\email{zhangxingh03@st.lzu.edu.cn}
\author{Yuxiao Liu}
\author{Li Zhao}
\affiliation{ Institute of Theoretical Physics, Lanzhou
University,  Lanzhou 730000, People's Republic of China}

\begin{abstract}
By making use of the decomposition of $U(1)$ gauge potential
theory and the $\phi$-mapping method, we propose that a charged
two-condensate Bose system possesses vortex lines and two classes
of knotted solitons. The topological charges of the vortex lines
are characterized by the Hopf indices and the Brower degrees of
$\phi$-mapping, and the knotted solitons are described by the
nontrivial Hopf invariant and the BF action, respectively.
\end{abstract}

\maketitle
%74.20.De Phenomenological theories (two-fluid, Ginzburg-Landau, etc.)
%11.15.Ex Spontaneous breaking of gauge symmetries
%03.75.Mn Multicomponent condensates; spinor condensates
%\section{introduction}

\section{Introduction}
Multigap superconductivity has been discussed in the late 1950's
for materials with a varying strength of electron-phonon
interactions between different pieces of the Fermi surface
\cite{mul1,mul2}. Since then, great attention has been directed
towards understanding the detailed nature of superconductivity in
this simple intermetallic compound. In recent years, \emph{ab
inito} calculations \cite{ab1,ab2,ab3} have showed that $MgB_2$
has two weakly coupled gaps: $\Delta_{\sigma}\approx7 mev$ and
$\Delta_{\pi}\approx2.5 mev$, residing on disconnected sheets of
the Fermi surface formed by in-plane \emph{$P_{xy}$} boron
orbitals (the $\sigma$-bands) and out-of -plane \emph{$P_{z}$}
boron orbitals (the $\pi$-bands). The most striking consequences
of the two gaps are the unusual anisotropic feature of $MgB_2$
under magnetic field, for example inequality between the
penetration depth and the upper critical field anisotropy, and
their variations with temperature \cite{European}. Moreover, the
experiment of scanning tunneling spectroscopy (STS) measurements
on single crystal $MgB_2$ shows that the coherent length in the
$\pi$ band is approximately 50 nm \cite{jiang11} which is much
larger than an estimate one that would obtain from a standard GL
formula. All these spark renewed interest to two-gap
superconductivity.

Principally, the two-gap superconductivity can be investigated in
the frame of a charged two-condensate Bose system \cite{jiang}.
This system is described by a Ginzburg-Landau (GL) model with two
flavors of Cooper pairs. Alternatively, it relates to a
Gross-Pitaevskii (GP) functional with two charged condensates of
tightly bound fermion pairs, or some other charged bosonic fields
\cite{f2002}. Such theoretical models have a wider range of
applications, including interference between two Bose condensates
\cite{science}, a multiband superconductor \cite{twoband},
two-component Bose-Einstein condensates \cite{twobec} and
superconducting gap structure of spin-triplet superconductor
$Sr_2RuO_4$ \cite{gap}.

In this paper we mainly focus attention on the charged
two-condensate Bose system. Since topological properties have
played an important roles in the condensed matter, we shall be
particularly interested in those of the charged two-condensate
Bose system. We know that in a single-condensate system, the best
known topological object is the Abrikosov vortex \cite{0601347}.
But the advent of the charged two-condensate Bose system has
opened a new possibility for us to have far more topological
objects. Based on an extended version of Faddeev's $O(3)$
nonlinear $\sigma$ model, Babaev \emph{et. al.} argued that the
knotted soliton, which is a stable, finite-length, closed vortex
line, exists in the charged two-condensate Bose system. The
knotted solitons are much more complex and structurally
complicated topological defects than Abrikosov vortices, and thus
its realization should open an exceptionally wide range of
possibilities of studies of various phenomena associated with them
\cite{prl2002}. Generally, the numerical simulations method was
adopted to research solitons \cite{jiang20}, but to our knowledge
there have been no real attempts to find an actual solution.
Nevertheless the $\phi$-mapping method \cite{phi1} provides a much
briefer method, by making use of which one needs not to solve the
nonlinear equation but directly investigates the knotted soliton
from the point of view of topology. In this paper, using the novel
topological current method, we study the topological properties of
the charged two-condensate Bose system in more detail and
naturally obtain two classes of knotted solitons, which are
described by a nontrivial Hopf invariant and the BF action,
respectively.

\section{Topological vortex lines in the two charged system}

A system with two electromagnetically coupled, oppositely charged
Bose condensates can be described by a two--flavor (denoted by
$\alpha=1, 2$) Ginzburg-Landau or Gross-Pitaevskii (GLGP)
functional, whose free energy density is given by
\begin{eqnarray}\label{F}
f&=&\frac{1}{2m_{1}}|(\hbar\partial_{\lambda}
+i\frac{2e}{c}A_{\lambda})\nonumber
\Psi_{1}|^{2}+\frac{1}{2m_{2}}|(\hbar\partial_{\lambda}
-i\frac{2e}{c}A_{\lambda})\Psi_{2}|^{2}\\
&&+V(\Psi_{1,2})+\frac{\textbf{B}^{2}}{8\pi}
\end{eqnarray}
where $V(\Psi_{1,2})$ is expressed as:
$V(\Psi_{1,2})=-b_\alpha|\Psi_\alpha|^2+\frac{c_\alpha}{2}|\Psi_\alpha|^4$.
The two condensates are characterized by the different effective
masses $m_{\alpha}$, the different coherence lengths
$\xi_\alpha\!\!=\!\!\hbar/\sqrt{2m_\alpha{b}_\alpha}$ and the
different concentrations
$N_\alpha=\langle|\Psi_\alpha|^2\rangle=b_\alpha/c_\alpha$. The
properties of the corresponding model with a single charged Bose
field are well known. In that case the field degrees of freedom
are the massive modulus of the single complex order parameter and
a vector field which gains a mass due to the Meissner-Higgs
effect. What is much important in the present GLGP model is that
the two charged fields are not independent but nontrivially
coupled through the electromagnetic field. This kind of nontrivial
coupling indicates that in this system there should be a
nontrivial, hidden topology which, however, cannot be recognized
obviously in the form of Eq. (\ref{F}). In order to find out the
topological structure and to investigate it conveniently, Ref.
\cite{f2002} introduce a set of new variables, involving a massive
field $\rho$ which is related to the densities of the Cooper pair
and a three-component unit vector field $\hat{n}$. Then the
original GLGP free energy density Eq. (\ref{F}) can be represented
as
\begin{eqnarray}\label{Fnew1}
f&=&\frac{\hbar^{2}\rho^{2}}{4}(\partial{n})^2+\hbar^{2}(\partial\rho)^{2}
+\frac{\hbar^{2}c^{2}}{512\pi{e}^{2}}\nonumber
\{\frac{1}{\hbar}[\partial_{\mu}C_{\nu}-\partial_{\nu}C_{\mu}]\\
&&-\hat{n}\cdot\partial_\mu\hat{n}\times\partial_\nu\hat{n}\}^{2}
+\frac{\rho^{2}}{16}\vec{C}^{2}+V.
\end{eqnarray}
where $\vec{C}=\vec{J}/{e\rho^{2}}$ and $\vec{J}$ stands for the
standard supercurrent. Now we can see that there allows an exact
equivalence between the two-flavor GLGP model and the nonlinear
$O(3)$ $\sigma$ model \cite{jiang20} which is much more important
to describe the topological structure in high energy physics. In
this paper, based on the decomposition of $U(1)$ gauge potential
theory and the $\phi$-mapping method, we discuss the counterpart
in condensed matter and show that there are topological
excitations in the form of stable knotted closed vortices in the
charged two-condensate Bose system. As shown in Eq. (\ref{Fnew1}),
we know that the magnetic field of the system can be divided into
two parts: One is the contribution of field $C_\mu$, we learn that
this part is introduced by the supercurrent density and can only
present us with the topological defects as what  in a
single-condensate system; The other part, the contribution
$\hat{n}\cdot\partial_\mu\hat{n}\times\partial_\nu\hat{n}$ to the
magnetic field term in Eq. (\ref{Fnew1}), is a fundamentally
topological property of the two-condensate system which has no
counterpart in the single-condensate system. Here we emphasize
that there allow another nontrivial topological configurations,
which are originated from the interaction between the field
$C_\mu$ and the contribution of the term
$\hat{n}\cdot\partial_\mu\hat{n}\times\partial_\nu\hat{n}$. Thus
we will investigate these terms in detail.

It is easy to prove that this term
$\hat{n}\cdot\partial_\mu\hat{n}\times\partial_\nu\hat{n}$ can be
reexpressed in an Abelian field tensor form \cite{U(1)current}: $
\hat{n}\cdot\partial_\mu\hat{n}\times\partial_\nu\hat{n}
=\partial_\mu{b}_\nu-\partial_\nu{b}_\mu$. Here $b_\mu$ is the
Wu-Yang potential \cite{wang15}
\begin{equation}\label{bu}
b_\mu=\hat{e}_1\cdot\partial_\mu\hat{e}_2,
\end{equation}
in which $\hat{e_1}$ and $\hat{e_2}$ are two perpendicular unit
vectors normal to $\hat{n}$, and $(\hat{e}_1,\hat{e}_2,\hat{n})$
forms an orthogonal frame:
$\hat{n}=\hat{e_1}\times{\hat{e_2}},\;\;
\hat{e}_1\cdot{\hat{e}_2}=0$. Now, consider a two-component vector
field $\vec{\phi}=(\phi^1,\phi^2)$ residing in the plane formed by
$\hat{e}_1$ and $\hat{e}_2$:
$e^a_1=\frac{\phi^a}{\|\phi\|},\;e^a_2=\epsilon_{ab}\frac{\phi^b}{\|\phi\|},\;
(\|\phi\|^2=\phi^a\phi^a;\; a,b=1,2)$. It can be proved that the
relations for $\hat{e}_1$ and $\hat{e}_2$ satisfy above
restriction. Obviously the zero points of
$\vec{\phi}=(\phi^1,\phi^2)$ are the singular points of
$\hat{e}_1$ and $\hat{e}_2$. We will find out that the topological
defects are just originated from these points. Using the
$\vec\phi$ field, the Wu-Yang potential Eq. (\ref{bu}) can be
expressed as
$b_\mu=\epsilon_{ab}\frac{\phi^a}{\|\phi\|}\partial_{\mu}\frac{\phi^b}{\|\phi\|}$.
Comparing with the expression of the $U(1)$ gauge potential
decomposition \cite{phi3},  we learn that $b_\mu$ satisfies the
$U(1)$ gauge transformation. The GLGP free energy density  Eq.
(\ref{Fnew1}) can be reexpressed as
\begin{eqnarray}\label{Fnew}
f&=&\frac{\hbar^{2}\rho^{2}}{4}(\partial{n})^2+\hbar^{2}(\partial\rho)^{2}\nonumber
+\frac{\hbar^{2}c^{2}}{512\pi{e}^{2}}\\
&&\times\{\frac{1}{\hbar}[\partial_{\mu}C_{\nu}-\partial_{\nu}C_{\mu}]
-H_{\mu\nu}\}^{2} +\frac{\rho^{2}}{16}\vec{C}^{2}+V,
\end{eqnarray}
where $H_{\mu\nu}$ is a two order tensor and can be expressed as:
$H_{\mu\nu}=2\epsilon_{ab}\partial_\mu(\phi^a/\|\phi\|)
\partial_\nu(\phi^b/\|\phi\|)$, which
describes the magnetic field that becomes induced in the system
due to a nontrivial electromagnetic interaction between the two
condensates. Indeed, it is exactly due to the presence of this
term that the two-condensate system acquires properties which are
qualitatively very different from those of a single-condensate
system.
%\section{knotted solitons in the system}\label{sec3}

Since this tensor $H_{\mu\nu}$ in Eq. (\ref{Fnew}) plays an
important role in the topological feature of the two-condensate
system, here we will use the $\phi$-mapping method to research the
properties hidden in this tensor. To do so, we introduce a
topological tensor current $K^{\mu\nu}$ \cite{phi3}, which is
denoted by
\begin{equation}\label{8}
K^{\mu\nu}=\frac{1}{8\pi}\epsilon^{\mu\nu\lambda\rho}H_{\lambda\rho}
=\frac{1}{4\pi}\epsilon^{\mu\nu\lambda\rho}\epsilon_{ab}\partial
_\lambda\frac{\phi^a}{\|\phi\|}\partial_\rho\frac{\phi^b}{\|\phi\|}.
\end{equation}
Using $\partial_{\mu}({\phi^a}/{\|\phi\|})=({\partial_\mu\phi^a})/
{\|\phi\|}+\phi^a\partial_\mu({1}/{\|\phi\|})$ and the Green
function relation in ${\phi}$ space:
$\partial_{a}\partial_{a}ln\|{\phi}\|=2\pi\delta^{2}({\vec{\phi}})\;
(\partial_{a}={\partial}/{\partial{\phi}^{a}})$, one can prove
that $K^{\mu\nu}=\delta^{2}({\vec{\phi}})D^{\mu\nu}({\phi}/{x})$,
where
$D^{\mu\nu}({\phi}/{x})=\frac{1}{2}\epsilon^{\mu\nu\lambda\rho}
\epsilon_{ab}\partial_{\lambda} \phi^{a}\partial_{\rho}{\phi}^{b}$
is the Jacobian between the $\phi$ space and the Cartesian
coordinates. Denoting the spacial components of $K^{\mu\nu}$ by
$K^i$, we obtain
\begin{equation}\label{K}
K^i=K^{0i}=\delta^{2}({\vec{\phi}})D^{i}(\frac{{\phi}}{x})\;\;(i=1,2,3),
\end{equation}
where $D^{i}(\frac{{\phi}}{x})=D^{0i}(\frac{{\phi}}{x})$. The
expression Eq. (\ref{K}) provides an important conclusion:
$K^i=0,\;\textrm{if and only if}\;{\vec{\phi}}\neq0$; $
K^i\neq0,\;\textrm{if and only if}\;{\vec{\phi}}=0$. So it is
necessary to study the zero points of ${\vec{\phi}}$ to determine
the nonzero solutions of $K^i$. The implicit function theory \cite
{implicit} shows that under the regular condition $D^i({\phi}
/x)\neq 0$, the general solutions of
\begin{equation}\label{phi}
{\phi}^1(t,x^1,x^2,x^3)=0,\;\;{\phi}^2(t,x^1,x^2,x^3)=0
\end{equation}
can be expressed as $\vec{x}=\vec{x}_k(s,t)$, which represent the
world surfaces of $N$ moving isolated singular strings $L_{k}$
with string parameter $s\;(k=1,2 \cdots N)$. This indicates that
in the charged two-condensate Bose system, there are vortex lines
located at the zero points of the $\vec{\phi}$ field.

Then question are raised naturally: what is the topological charge
of the vortex lines and how to obtain the inner topological
structure of $K^i$. Now we will investigate the topological
charges of the vortex lines and their quantization. In
$\delta$-function theory \cite{delta}, one can prove that in
three-dimensional space
\begin{equation}\label{delta}
\delta^{2}(\vec{\phi})=\sum^{N}_{k=1}\beta_{k}\int_{L_{k}}
\frac{\delta^{3}(\vec{x}-\vec{x}_{k}(s))}{|D(\frac{\phi}{u})|_{\Sigma_{k}}}ds,
\end{equation}
where $ D(\phi/u)_{\Sigma_{k}}=
\frac{1}{2}\epsilon^{\mu\nu}\epsilon_{mn}
(\partial\phi^{m}/\partial{u^{\mu}})(\partial\phi^{n}/\partial{u^{\nu}})$
and $\Sigma_{k}$ is the $k$th planar element transverse to $L_{k}$
with local coordinates $(u^{1}, u^{2})$. The positive integer
$\beta_{k}$ is the Hopf index of the $\phi$-mapping, which means
that when $\vec{x}$ covers the neighborhood of the zero point
$\vec{x}_{k}(s)$ once, the vector field $\vec\phi$ covers the
corresponding region in $\phi$ space $\beta_{k}$ times. Meanwhile
taking notice of the definition of the Jacobian, the direction
vector of $L_{k}$ is given by
\begin{equation}\label{volicity}
\left.\frac{dx^{i}}{ds}\right|_{\vec{x}_{k}}=\left.\frac{D^{i}
(\phi/x)}{D(\phi/u)}\right|_{\vec{x}_{k}}.
\end{equation}
Then from Eqs. (\ref{delta}) and (\ref{volicity}), we obtain the
inner structure of $K^i$
\begin{equation}\label{ji2}
K^{i}=\delta^{2}(\vec{\phi})D^{i}(\frac{\phi}{x})
=\sum^{N}_{k=1}\beta_k\eta_k\int_{L_{k}}\frac{dx^{i}}{ds}
\delta^{3}(\vec{x}-\vec{x}_{k}(s))ds,
\end{equation}
in which $\eta_{k}$ is the Brouwer degree of the $\phi$-mapping,
with $\eta_{k}=\textrm{sgn}D(\phi/u)_{\Sigma_{k}}=\pm1$. Hence the
topological charge of the vortex line $L_{k}$ is
$Q_{k}=\int_{\Sigma_{k}}K^{i}d\sigma_{i}=W_{k}$, where
$W_{k}=\beta_{k}\eta_{k}$ is the winding number of $\vec{\phi}$
around $L_k$. Then we come to the conclusion: under the regular
condition $D^i(\phi/x)\neq0$ there exist vortex lines in the
charged two-condensate Bose system, whose topological charges are
just the winding numbers of the $\phi$-mapping.

\section{topological properties of knotted solitons}

In this section, we research stationary knots corresponding to the
vortex lines in this two-condensate system. When going to a
constant vector $\hat{n}(x)\rightarrow{\hat{n}_0}$, $\hat{n}(x)$
naturally defines a mapping: $R_3\thicksim{S^3}\rightarrow{S^2}$
\cite{jiang20}. Such mappings fall into nontrivial homotopy
classes $\pi_3(S^2)\thickapprox{Z}$ and can be characterized by a
topological invariant called the Hopf invariant
\cite{nature3,nature4}, an integral formula which counts the
number of times the two-sphere $S^2$ is covered by the
three-sphere $S^3$. From the above discussion, the Hopf invariant
can be given in terms of $b_\mu$ and $H_{\mu\nu}$ by
\cite{jiang20}
\begin{equation}\label{action1}
I=\frac{1}{16\pi^2}\int\epsilon^{ijk}{b}_iH_{jk}d^3x,
\end{equation}
where $b_i$ and $H_{jk}$ are the spacial components of $b_\mu$ and
$H_{\mu\nu}$, respectively. It can be seen that when these $N$
vortex lines are closed curves, i.e., a family of knots $\gamma_k$
$(k=1,\cdots,N)$, considering Eq. (\ref{8}) and substituting Eq.
(\ref{ji2}) into Eq. (\ref{action1}), we obtain
\begin{equation}\label{action3}
I=\frac{1}{2\pi}\sum_{k=1}^NW_k\oint_{\gamma_k}b_i{dx^i}.
\end{equation}
This is a very important expression. Consider the $U(1)$ gauge
transformation of $b_\mu$: $b_\mu'=b_\mu+\partial_i\theta$, where
$\theta\in{I\!\!R}$ is a phase factor denoting the $U(1)$
transformation. It is seen that the term $\partial_i\theta$
contributes noting to the integral $I$, hence the expression
(\ref{action3}) is invariant under the $U(1)$ gauge
transformation. Meanwhile we know that $I$ is independent of the
metric $g_{\mu\nu}$. Therefore one can conclude that $I$ is a
topological invariant for knotted vortex lines in the charged
two-condensate Bose system. In the presence of the vector field
$\hat{n}$, the two-condensate system allows this kind of knots
labelled by the Hopf invariant.

Simultaneously, it is due to the interaction between the massive
field $C_\mu$ and the tensor $H_{\mu\nu}$ that a new nontrivially
topological configuration will be produced in the two-condensate
system. One can see that $C=C_\mu{dx^\mu}$ is a one-form, which
presents us with the properties as what in a single-condensate
system, and the two-form
$H=\frac{1}{2}H_{\mu\nu}dx^\mu\wedge{dx}^\nu$ describes the
magnetic field that becomes induced to the interaction between the
two condensate. In order to explore the new topological
configuration according to the interaction between the one-form
$C$ and the two-form $H$, we introduce the classical Abelian BF
theory \cite{BF,JHEP}. In this three-dimensional charged
two-condensate system, the abelian BF theory has the following
action
\begin{equation}\label{s}
S=\frac{1}{4\pi}\int_MC\wedge{H}=\frac{1}{8\pi}\int_M\epsilon^{ijk}C_iH_{jk}dx^3,
\end{equation}
where $C_i$ and $H_{jk}$ are the spacial components of $C_\mu$ and
$H_{\mu\nu}$. Considering the relation
$K^i=\frac{1}{8\pi}\epsilon^{ijk}H_{jk}$, one knows that there
also allow the vortex lines. When the vortex lines are a family of
knots, substituting Eq. ({\ref{ji2}}) into Eq. (\ref{s}), one can
obtain
\begin{equation}\label{BF}
S=\sum^N_{f=1}W_f\oint_{\gamma_{f}}C_idx^i.
\end{equation}
We learn that $C_\mu$ satisfies the $U(1)$ gauge transformation:
$C_\mu'=C_\mu+\partial_\mu\varphi$ and $\varphi\in{I\!\!R}$ is a
phase factor denoting the $U(1)$ transformation. Furthermore the
term $\partial_\mu\varphi$ contributes noting to the integral $S$,
hence the expression (\ref{BF}) is invariant under the $U(1)$
gauge transformation. Meanwhile one knows that $S$ is independent
of the metric $g_{\mu\nu}$. Therefore we can conclude that the BF
action $S$ is also a topological invariant for knotted vortex
lines in the charged two-condensate Bose system.

It is well known that many important topological numbers are
related to a knot family such as the self-linking number and Gauss
linking number. Next we will discuss the relationship between the
actions Eqs. (\ref{action3},\ref{BF}) and the topological numbers
of the knot family. First $b_i$ and $C_i$ should be expressed in
terms of the vector field which carries the geometric information
of the knot family. Define the Gauss mapping
$\vec{m}:S^1\times{S^1}\rightarrow{S^2}$ by
$\vec{m}(\vec{y},\vec{x})=\frac{\vec{y}-\vec{x}}{||\vec{y}-\vec{x}||}$,
where $\vec{x}$ and $\vec{y}$ are two points, respectively, on the
knots $\gamma_{k}$ and $\gamma_{l}$. Let a two-dimensional unit
vector $\vec{u}=\vec{u}\;(\vec{x},\vec{y})$ satisfy
$u^{a}u^{a}=1$, where $a=1,2\;\textrm{and}\; \vec{u}\perp\vec{m}$.
Then $b_i$ can be decomposed in terms of this two-dimensional
vector $u^{a}$:
$b_i=\epsilon^{ab}u^{a}\partial_{i}u^{b}-\partial_{i}\alpha$, in
which $\alpha$ is a phase factor. it is due to the term
$\partial_{i}\alpha$ contributes nothing to the integral Eq.
(\ref{action3}) that $b_i$ can be written as
$b_i=\epsilon^{ab}u^{a}\partial_{i}u^{b}$.
%$C_i=\epsilon^{ab}u^{a}\partial_{i}u^{b}$.
And noticing the symmetry between the point $\vec{x}$ and
$\vec{y}$, Eq. (\ref{action3}) can be reformulated as
\begin{eqnarray}\label{actioni}
I&=&\frac{1}{2\pi}\sum^{N}_{k=1}\sum^{N}_{l=1}W_{k}W_{l}\\
&\times&\oint_{\gamma_{k}}\oint_{\gamma_{l}}
\epsilon_{ab}\partial_{i}u^{a}(\vec{x},\vec{y})\nonumber
\partial_{j}u^{b}(\vec{x},\vec{y})dx^{i}\wedge{dy^{j}}.
\end{eqnarray}

Further, we express the abelian BF action $S$ as a new form
including a unit vector, which also carries geometric information.
Introduce another Gauss mapping $\vec{d}:
S^1\times{S^1}\rightarrow{S^2}$ and a two-dimensional unit vector
field $\vec{v}(\vec{x'},\vec{y'})$, where $\vec{x'},\vec{y'}$ are
different points on $\gamma_f,\gamma_g$, and let $\vec{v},\vec{d}$
satisfy $\vec{v}\bot\vec{d}$. Then using the $U(1)$ gauge
potential decomposition theory, one can obtain that $C_i$ is
decomposed in terms of the vector $\vec{v}$:
$C_i=\epsilon^{ab}v^{a}\partial_{i}v^{b}+\partial_i\beta$, where
$\beta$ is a phase factor and $\partial_i\beta$ has no
contribution to the integral Eq. (\ref{BF}). Substituting the
expression of the decomposition into the BF action Eq. (\ref{BF})
and taking notice of the symmetry between the point $\vec{x'}$ and
$\vec{y'}$, we get
\begin{eqnarray}\label{actions}
S&=&\!\!\sum^{N}_{f=1}\sum^N_{g=1}\!\!W_{f}W_{g}\\\nonumber
&\times&\oint_{\gamma_{f}}\oint_{\gamma_{g}}
\epsilon_{ab}\partial_{i}v^{a}(\vec{x'},\vec{y'})
\partial_{j}v^{b}(\vec{x'},\vec{y'})dx'^{i}\wedge{dy'^{j}}.
\end{eqnarray}
In fact there exist three cases for Hopf invariant Eq.
(\ref{actioni}) and the BF action Eq. (\ref{actions}),
respectively. (i) The same knots but the different points, i.e.,
$k=l,\vec{x}\neq\vec{y}$ or $f=g, \vec{x'}\neq\vec{y'}$; (ii) The
same knots and the same points, i.e., $k=l,\vec{x}=\vec{y}$ or
$f=g,\vec{x'}=\vec{y'}$; (iii) The different knots, i.e.,
$k\neq{l}$ or $f\neq{g}$. For the first case, when
$\gamma_k,\gamma_l$ are the same knots, $\gamma_f,\gamma_g$ are
the same knots, but $\vec{x}$ and $\vec{y}$, $\vec{x'}$ and
$\vec{y'}$ are different points, taking account of the relations
$\epsilon_{ab}\partial_iu^a\partial_ju^b=\frac{1}{2}\vec{m}
\cdot(\partial_i\vec{m}\times\partial_j\vec{m})$ and
$\epsilon_{ab}\partial_iv^a\partial_jv^b=\frac{1}{2}\vec{d}
\cdot(\partial_i\vec{d}\times\partial_j\vec{d})$ , one can prove
that
%\begin{widetext}
\begin{eqnarray}
&&\frac{1}{2\pi}\oint_{\gamma_{k}}\oint_{\gamma_{k}}
\epsilon_{ab}\partial_{i}u^{a}\partial_{j}u^{b}dx^{i}\wedge{dy^{j}}
=\frac{1}{4\pi}\oint_{\gamma_k}\oint_{\gamma_k}\vec{m}^*(dS)\nonumber\\
&&\frac{1}{2\pi}\oint_{\gamma_f}\oint_{\gamma_f}\nonumber
\epsilon_{ab}\partial_{i}v^{a}\partial_{j}v^{b}dx'^{i}\wedge{dy'^{j}}
=\frac{1}{4\pi}\oint_{\gamma_f}\oint_{\gamma_f}\vec{d}^*(dS)
\end{eqnarray}
where $\vec{m}^*(dS)=\vec{m}\cdot(d\vec{m}\times
{d\vec{m}})=2\epsilon_{ab}du^a\wedge{du^b}$ is the pull-back of
$S^2$ surface element. The expressions is just related to the
writhing number $Wr(\gamma_k)$ of $\gamma_k$:
$W(\gamma_k)=\frac{1}{4\pi}\oint_{\gamma_k}\oint_{\gamma_k}\vec{m}^*(dS)$.
For the second case, $\gamma_k$, $\gamma_l$ are the same knots,
$\gamma_f$, $\gamma_g$ are the same, $\vec{x}$ and $\vec{y}$ are
just the same points, $\vec{x'}$ and $\vec{y'}$ are also the same.
Let $\vec{T}$ ($\vec{T'}$) be the unit tangent vector of knot
$\gamma_k$ ($\gamma_f$) at $\vec{x}$ ($\vec{x'}$) and let
$\vec{V}$ ($\vec{V'}$) satisfy $u^a=\epsilon^{ab}V^b$
($v^a=\epsilon^{ab}V'^b)$, where
$a,b=1,2;\;\vec{V}\bot\vec{T},\;\vec{u}=\vec{T}\times\vec{V}$
($\vec{V'}\bot\vec{T'},\;\vec{v}=\vec{T'}\times\vec{V'}$), then it
is easy to prove
\begin{eqnarray}\nonumber
&&\frac{1}{2\pi}\oint_{\gamma_k}\epsilon_{ab}u^a\partial_iu^bdx^i=\nonumber
\frac{1}{2\pi}\oint_{\gamma_k}(\vec{T}\times\vec{V})\cdot{d\vec{V}}=Tw(\gamma_k),\\
&&\frac{1}{2\pi}\oint_{\gamma_f}\epsilon_{ab}v^a\partial_iv^bdx'^i=\nonumber
\frac{1}{2\pi}\oint_{\gamma_f}(\vec{T'}\times\vec{V'})\cdot{d\vec{V'}}=Tw(\gamma_f),
\end{eqnarray}
where $Tw(\gamma_k)$ and $Tw(\gamma_f)$ are the twisting numbers
of the knot $\gamma_k$ and $\gamma_f$. According to the
Calugareanu's formula  \cite{white1, white2}:
$SL(\gamma_k)=Wr(\gamma_k)+Tw(\gamma_k)$, where $SL(\gamma_k)$ is
just the self-linking number of $\gamma_k$, one can arrive at the
result that when $\gamma_k$ and $\gamma_l$, $\gamma_f$ and
$\gamma_g$ are the same knots i.e. $k=l,f=g$, the actions Eqs.
(\ref{actioni}) and (\ref{actions}) are related to the
self-linking of knots. For the third case, when $\gamma_k$ and
$\gamma_l$, $\gamma_f$ and $\gamma_g$ are different knots,
considering the relations between $\vec{u}$ and $\vec{m}$,
$\vec{v}$ and $\vec{d}$, we can obtain
\begin{eqnarray}
\frac{1}{2\pi}\oint_{\gamma_{k}}\oint_{\gamma_{l}}
\epsilon_{ab}\partial_{i}u^{a}
\partial_{j}u^{b}dx^{i}\wedge{dy^{j}}\nonumber
=\frac{1}{4\pi}\oint_{\gamma_k}\oint_{\gamma_l}\vec{m}^*(dS)\\
\frac{1}{2\pi}\oint_{\gamma_{f}}\oint_{\gamma_{g}}
\epsilon_{ab}\partial_{i}v^{a}\nonumber
\partial_{j}v^{b}dx'^{i}\wedge{dy'^{j}}
=\frac{1}{4\pi}\oint_{\gamma_f}\oint_{\gamma_g}\vec{d}^*(dS),
\end{eqnarray}
in which
$\frac{1}{4\pi}\oint_{\gamma_k}\oint_{\gamma_l}\vec{m}^*(dS)\!\!=\!\!\frac{1}{4\pi}\epsilon^{ijk}
\oint_{\gamma_{k}}dx^i\oint_{\gamma_{l}}
dy^i\frac{(x^k-y^k)}{\|\vec{x}-\vec{y}\|^3}\!\!\!=\!\!\!L(\gamma_k,\gamma_l)$
is the Gauss linking number between $\gamma_k$ and $\gamma_l$,
$\frac{1}{4\pi}\oint_{\gamma_f}\oint_{\gamma_g}\vec{d}^*(dS)
=L(\gamma_f,\gamma_g)$ is the Gauss linking number between
$\gamma_f$ and $\gamma_g$. Therefore, taking account of all these
three cases, we come to the important conclusions
\begin{eqnarray}
I&=&\sum^N_{k=1}\sum^N_{l=1}W_kW_lL(\gamma_k,\gamma_l)+\sum^N_{k=1}
 W^2_kSL(\gamma_k),\\
S&=&2\pi\left[\sum^N_{f=1}\sum^N_{g=1}W_fW_gL(\gamma_f,\gamma_g)+\sum^N_{f=1}
 W^2_fSL(\gamma_f)\right],\nonumber
\end{eqnarray}
which describe the twisted knot topology of the physical magnetic
flux. Since the self-linking number and the Gauss linking number
are both the invariant characteristic numbers of the knotted
closed curves in topology, $I$ and $S$ are also important
topological invariants required to describe the knotted vortex
lines in the charged two-condensate system. Notice that they can
be understood as the linking of two quantized magnetic fluxes of
the knotted vortex lines. Obviously two flux rings linked together
can not be separated by any continuous deformation of the field
configuration. This provides the topological stability of the
knots. Since the energy of a vortex line is proportional to its
length, for a finite energy the length must be finite
\cite{jiang20}. Then we obtain a stable, finite length closed
vortex lines in the charged two-condensate Bose system, namely,
knotted solitions. This is just the significance of the
introduction and research of topological invariants $I$ and $S$.

\section{Conclusion}
In conclusion, based on the decomposition of $U(1)$ gauge
potential theory and the $\phi$-mapping topological current
method, we point out there allow vortex lines and two classes of
knotted solitons in the charged two-condensate Bose system. Under
the regular condition $D^i(\phi/x)\neq0$, vortex lines are
originated form the zero points of the $\vec\phi$ field and the
topological charges of vortex lines are characterized by the
winding numbers of the $\phi$-mapping. Furthermore, we discuss the
stationary knots and find out two classes of knotted
solitons--stable, finite length, closed vortex lines in the
two-condensate system. One knotted solition is described by the
Hopf invariant and the other is characterized by the BF action.
Finally, the relations between these actions and the topological
numbers of a family knot are investigated in the system. It is
revealed that these actions are topological invariants for knotted
solitons and can be reformulated into the self-linking numbers and
the Gauss linking numbers.

At last, it should be pointed out that in the present paper we
treat the vortex lines as mathematical lines, i.e., the width of a
line is zero. This description is obtained in the approximation
that the radius of curvatures of a line is much larger than the
width of the line \cite{nielsen}.

\section*{Acknowledgments}
It is a great pleasure to thank Dr. P. M. Zhang for numerous
fruitful discussions. This work was supported by the National
Natural Science Foundation of China under Grant No. 10475034.
% and the Doctoral Foundation of the People's Republic of China.

\end{document}